\documentclass{aastex}
\usepackage{amsmath}
\usepackage{graphicx}
\usepackage{url}
\usepackage{amssymb}
\usepackage{float}
\makeatletter
\let\@dates\relax
\makeatother

\shorttitle{Albedo protons measured by PAMELA}
\shortauthors{Adriani et al.}

\begin{document}


\title{Re-Entrant Albedo Proton Fluxes Measured by the PAMELA Experiment}

\author{
O.~Adriani$^{1,2}$,
G.~C.~Barbarino$^{3,4}$,
G.~A.~Bazilevskaya$^{5}$,
R.~Bellotti$^{6,7}$,
M.~Boezio$^{8}$,
E.~A.~Bogomolov$^{9}$,
M.~Bongi$^{1,2}$,
V.~Bonvicini$^{8}$,
S.~Bottai$^{2}$,
A.~Bruno$^{6,*}$,
F.~Cafagna$^{7}$,
D.~Campana$^{4}$,
P.~Carlson$^{10}$,
M.~Casolino$^{11,12}$,
G.~Castellini$^{13}$,
C.~De~Donato$^{11,15}$,
C.~De~Santis$^{11,15}$,
N.~De~Simone$^{11}$,
V.~Di~Felice$^{11,16}$,
V.~Formato$^{8,17}$,
A.~M.~Galper$^{14}$,
A.~V.~Karelin$^{14}$,
S.~V.~Koldashov$^{14}$,
S.~Koldobskiy$^{14}$,
S.~Y.~Krutkov$^{9}$,
A.~N.~Kvashnin$^{5}$,
A.~Leonov$^{14}$,
V.~Malakhov$^{14}$,
L.~Marcelli$^{11,15}$,
M.~Martucci$^{15,18}$,
A.~G.~Mayorov$^{14}$,
W.~Menn$^{19}$,
M.~Merg$\acute{e}$$^{11,15}$,
V.~V.~Mikhailov$^{14}$,
E.~Mocchiutti$^{8}$,
A.~Monaco$^{6,7}$,
N.~Mori$^{1,2}$,
R.~Munini$^{8,17}$,
G.~Osteria$^{4}$,
F.~Palma$^{11,15}$,
B.~Panico$^{4}$,
P.~Papini$^{2}$,
M.~Pearce$^{10}$,
P.~Picozza$^{11,15}$,
M.~Ricci$^{18}$,
S.~B.~Ricciarini$^{2,13}$,
R.~Sarkar$^{20,21}$,
V.~Scotti$^{3,4}$,
M.~Simon$^{19}$,
R.~Sparvoli$^{11,15}$,
P.~Spillantini$^{1,2}$,
Y.~I.~Stozhkov$^{5}$,
A.~Vacchi$^{8}$,
E.~Vannuccini$^{2}$,
G.~I.~Vasilyev$^{9}$,
S.~A.~Voronov$^{14}$,
Y.~T.~Yurkin$^{14}$,
G.~Zampa$^{8}$
and N.~Zampa$^{8}$
}

\affil{$^{1}$ Department of Physics and Astronomy, University of Florence, I-50019 Sesto Fiorentino, Florence, Italy.}
\affil{$^{2}$ INFN, Sezione di Florence, I-50019 Sesto Fiorentino, Florence, Italy.}
\affil{$^{3}$ Department of Physics, University of Naples ``Federico II'', I-80126 Naples, Italy.}
\affil{$^{4}$ INFN, Sezione di Naples, I-80126 Naples, Italy.}
\affil{$^{5}$ Lebedev Physical Institute, RU-119991 Moscow, Russia}
\affil{$^{6}$ Department of Physics, University of Bari, I-70126 Bari, Italy.}
\affil{$^{7}$ INFN, Sezione di Bari, I-70126 Bari, Italy.}
\affil{$^{8}$ INFN, Sezione di Trieste, I-34149 Trieste, Italy.}
\affil{$^{9}$ Ioffe Physical Technical Institute, RU-194021 St. Petersburg, Russia.}
\affil{$^{10}$ KTH, Department of Physics, and the Oskar Klein Centre for Cosmoparticle Physics, AlbaNova University Centre, SE-10691 Stockholm, Sweden.}
\affil{$^{11}$ INFN, Sezione di Rome ``Tor Vergata'', I-00133 Rome, Italy.}
\affil{$^{12}$ RIKEN, Advanced Science Institute, Wako-shi, Saitama, Japan.}
\affil{$^{13}$ IFAC, I-50019 Sesto Fiorentino, Florence, Italy.}
\affil{$^{14}$ National Research Nuclear University MEPhI, RU-115409 Moscow, Russia.}
\affil{$^{15}$ Department of Physics, University of Rome ``Tor Vergata'', I-00133 Rome, Italy.}
\affil{$^{16}$ Agenzia Spaziale Italiana (ASI) Science Data Center, Via del Politecnico snc, I-00133 Rome, Italy.}
\affil{$^{17}$ Department of Physics, University of Trieste, I-34147 Trieste, Italy.}
\affil{$^{18}$ INFN, Laboratori Nazionali di Frascati, Via Enrico Fermi 40, I-00044 Frascati, Italy.}
\affil{$^{19}$ Department of Physics, Universit\"{a}t Siegen, D-57068 Siegen, Germany.}
\affil{$^{20}$ Indian Centre for Space Physics, 43 Chalantika, Garia Station Road, Kolkata 700084, West Bengal, India.}
\affil{$^{21}$ Previously at INFN, Sezione di Trieste, I-34149 Trieste, Italy.}
\affil{* Corresponding author. E-mail address: alessandro.bruno@ba.infn.it.}

\begin{abstract}
We present a precise measurement of down\-ward-going albedo proton fluxes for kinetic energy above $\sim$ 70 MeV performed by the PAMELA experiment at
an altitude between 350 and 610 km.
On the basis of a trajectory tracing simulation, the analyzed protons were classified into quasi-trapped, concentrating in the magnetic equatorial region, and un-trapped spreading over all latitudes, including both short-lived (precipitating) and long-lived (pseudo-trapped) components.
In addition, features of the penumbra region around the geomagnetic cutoff were investigated in detail.
PA\-ME\-LA results significantly improve the characterization of the high energy albedo proton populations at low Earth orbits.
\end{abstract}


\section{Introduction}\label{Introduction}

The Cosmic Ray (CR) albedo radiation has been detected since late 1940s by rockets and balloon experiments \citep{VanAllen,Ormes}. These particles, generated with upward-going directions by the interaction of CRs from the interplanetary space with the Earth's atmosphere,
are typically classified into ``splash'' and ``re-entrant'' albedo components, depending on the type of trajectories followed from the production site \citep{Treiman}: while the former refers to particles which are able to escape from the magnetosphere (i.e. allowed St\"{o}rmer trajectories), the latter includes particles whose trajectories are bent by the geomagnetic field back to the Earth (i.e. forbidden trajectories).

The re-entrant albedo (hereafter albedo) population comprises quasi-trapped and un-trapped components.
The former concentrates in the near equatorial region, inside and below the inner Van Allen belt \citep{Moritz,Hovestadt,AMS01}, and presents features similar to those of stably-trapped
particles in the radiation belts although it is characterized by limited lifetimes and much less intense fluxes.
The latter spreads over all latitudes \citep{AMS01b,Bidoli} including the so-called penumbra \citep{STORMER,SHEA}, a region around the local geomagnetic cutoff rigidity where particle trajectories
become chaotic and both allowed and forbidden bands of CR particle access are present (see \citealt{Cooke} for definitions of CR cutoffs).

New accurate measurements of the high energy ($>$ 70 MeV) CR radiation at low Earth orbits have been reported by the PAMELA mission \citep{PHYSICSREPORTS}. Thanks to the orbit and the high identification capabilities, the instrument is able to provide detailed information about particle fluxes in different regions of the terrestrial magnetosphere, including energy spectra, spatial and angular distributions.
PAMELA measurement of geomagnetically trapped protons, and quasi-trapped electrons and positrons can be found in publications \citep{PAMTRAPPED,QUASITRAPPED}. In addition, for the first time PAMELA has revealed the existence of a trapped antiproton component in the inner belt \citep{PAMELAtrappedpbars}. In this article the measurement of the albedo proton fluxes is presented.

\section{PAMELA Data Analysis}\label{PAMELA data analysis}

PAMELA is a space-based experiment designed for a precise measurement of the charged cosmic radiation in the kinetic energy range from some tens of MeV up to several hundreds of GeV \citep{Picozza}. Details about apparatus performance, particle selection and selection efficiencies evaluation can be found elsewhere \citep{ProHe,SOLARMOD}. The Resurs-DK1 satellite, which hosts the apparatus, has a semi-polar (70 deg inclination) and elliptical (350 $\div$ 610 km altitude) orbit, and it is three-axis stabilized.
The spacecraft orientation is calculated by an on board processor with an accuracy better than 1 deg which, together with the good angular resolution ($<$ 2 deg) of the PAMELA tracking system, allows particle direction to be measured with high precision. The PAMELA major axis is mostly oriented toward the zenith direction.

\subsection{Data Set}\label{Data set}
The analyzed data set includes protons acquired by PAMELA between 2006 July and 2009 September, corresponding to
a total live time of about 800 days.
Data registered between 2006 December 6 and 2007 January 10, characterized by high geomagnetic disturbance levels associated with the December 6, 13 and 14 solar particle events \citep{SEP2006}, were excluded from the sample.
The time variation of the PAMELA detector performance, mostly due to the sudden failure of some front-end chips in the tracking system,
were carefully studied and accounted for in the efficiencies estimate.

\subsection{Geomagnetic Models}\label{Geomagnetic models}
The IGRF10 \citep{IGRF10} and the TS05 \citep{TS05} models were used for the description of internal and external geomagnetic field sources, respectively:
the former is a spherical harmonic expansion of the Earth's main magnetic field; the latter is able to provide a dynamical description (with a five-minute resolution) of the
geomagnetic field in the inner magnetosphere, based on recent satellite measurements.
Solar wind and interplanetary magnetic field values were derived from the high resolution Omniweb database \citep{OMNIWEB}.
All parameters were evaluated on an event-by-event basis through linear interpolation.

\subsection{Coordinate Systems}\label{Coordinate systems}
Data were analyzed in terms of ``Altitude Adjusted Corrected GeoMagnetic'' (AACGM) coordinates, developed to provide a more realistic description of high latitude regions by accounting for the multipolar geomagnetic field. They are defined such that all points along a magnetic field line have the same geomagnetic latitude and longitude, so that they are
closely related to invariant magnetic coordinates \citep{Baker, Gustafson, Heres}.
The AA\-CGM reference frame coincides with the standard ``Corrected GeoMagnetic'' (CGM) coordinate system \citep{BREKKE} at the Earth's surface.

\begin{figure*}[t]
\centering
\includegraphics[width=6.6in]{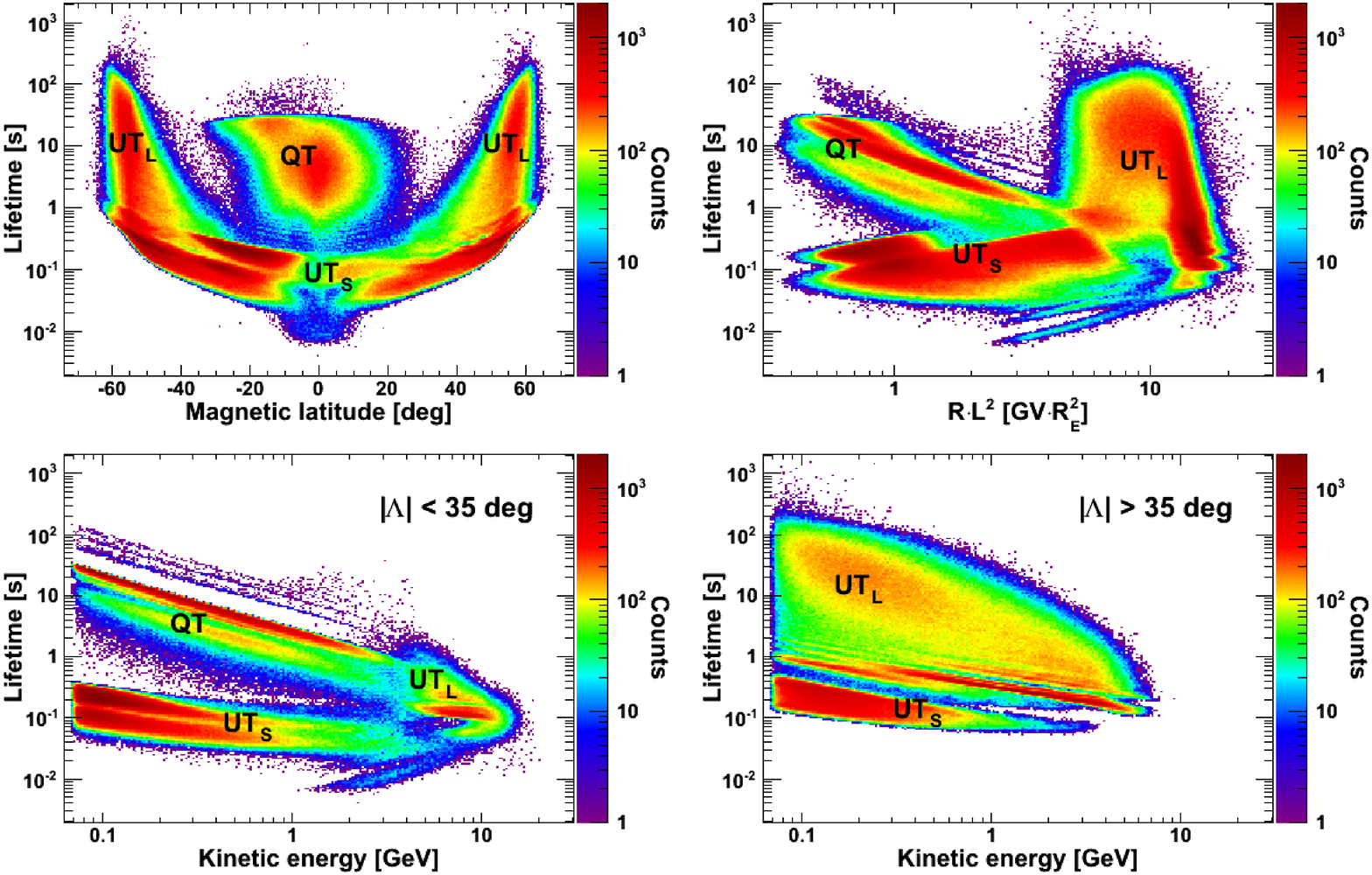}
\caption{Top panels: lifetime distributions for the measured sample (counts) as a function of the AACGM latitude $\Lambda$ (left) and of the product of particle rigidity $R$ and $L$-shell squared (right). Bottom panels: distributions as a function of kinetic energy, for AACGM latitudes $|\Lambda|$ lower and greater than 35 deg (left and right, respectively). The several albedo components can be easily distinguished: quasi-trapped (QT), precipitating (UT$_{S}$) and pseudo-trapped (UT$_{L}$).}
\label{Figure1}
\end{figure*}

\section{Trajectory Reconstruction}\label{Trajectory reconstruction}
Using spacecraft ephemeris data (position, orientation, time), and the particle rigidity ($R$ = momentum/charge) and direction provided by the magnetic spectrometer,
trajectories of all selected protons were reconstructed in the Earth's magnetosphere by means of a tracing program based on numerical integration methods \citep{TJPROG,SMART}, implementing the afo\-re\-men\-tio\-ned geomagnetic field models,
and carefully optimized for the calculation of very long trajectories in the magnetosphere.

Measured rigidities were corrected for the mean energy loss in the apparatus, evaluated with realistic simulations of the instrument.
This solution was preferred to standard statistical methods (e.g. fluxes unfolding) in order to provide a more reliable input to the trajectory simulation,
and to preserve spectral structures especially near the penumbra region.
The effects of the finite rigidity and angular resolution of PAMELA on the trajectory reconstruction were accounted for in the estimate of systematic uncertainties.

\subsection{Proton Classification}\label{Proton classification}
Trajectories were propagated back and forth from the measurement location, and
traced until:
\begin{enumerate}
\item they reached the model magnetosphere boundaries;
\item or they intersected the absorbing atmosphere limit, which was assumed at an altitude of 40 km (mean albedo production altitude);
\item or they performed more than $\sim$ $10^{6}/R^{2}$ steps for both propagation directions, where $R$ is the particle rigidity in GV.
\end{enumerate}
Particles satisfying the first condition were classified as protons from interplanetary space.
The second case corresponds to albedo protons.
Finally, the third category includes stably-trapped protons from the inner Van Allen belt, detected in the so-called ``South Atlantic Anomaly'' (SAA) region at PAMELA altitudes.
Since the program uses a dynamic variable step length, which is of the order of 1\% of a particle gyro-distance in the magnetic field,
the applied rigidity-dependent criterion (3) ensures that at least four drift cycles around the Earth were performed;
the measurement of such a component is described elsewhere \citep{PAMTRAPPED}.
The initial sample also included a negligible fraction of escaping albedo protons with rigidity above the geomagnetic cutoff.

Each albedo trajectory was checked by considering, as a first approximation, the particle motion in the magnetosphere as a superposition of three quasi-periodic motions: a gyration around field lines, a bounce between magnetic mirrors in opposite hemispheres, and an azimuthal drift around the Earth.
By using combined selections on corresponding mean frequencies ($\omega_{gyro}$, $\omega_{bounce}$ and $\omega_{drift}$)
as a function of several variables of interest (energy, pitch angle, 
latitude, etc.), the selected sample was sub-divided into two main categories:
\begin{itemize}
\item events with trajectories similar to those of stably-trapped protons,
    but originated and re-absorbed by the atmosphere during a time shorter than a few drift periods, were identified as quasi-trapped (QT). Their trajectories were verified to satisfy the adiabatic conditions, in particular the hierarchy of temporal scales: $\omega_{bounce}/\omega_{gyro}$ $\lesssim$ $\epsilon_{1}$ and $\omega_{drift}/\omega_{bounce}$ $\lesssim$ $\epsilon_{2}$, where
    $\epsilon_{1}$ and $\epsilon_{2}$ were estimated to be of the
    order of $10^{-2}\div10^{-1}$.
\item The rest of the sample was classified as un-trapped (UT). Qualitatively, two sub-components can be identified:
\begin{itemize}
\item precipitating protons (UT$_{S}$), with lifetimes short\-er than a bounce period. Corresponding $\omega_{bounce}$ values are similar to those of quasi-trapped protons, while $\omega_{gyro}$ distribution is much broader outside the SAA, extending to much lower values.
\item Pseudo-trapped protons (UT$_{L}$), with relatively long lifetimes. They are characterized by large gyro-radii and $\omega_{drift}$, and by small $\omega_{gyro}$ and $\omega_{bounce}$ values, resulting in unstable trajectories due to resonances occurring between component frequencies.
    They can perform several drift cycles (up to a few hundreds) reaching large distances from the Earth's surface, sometimes forming intermediate loops, before they are re-ab\-sor\-bed by the atmosphere.
\end{itemize}
\end{itemize}

\subsection{Lifetime Distributions}\label{Lifetime distributions}
The lifetime of the different populations was estimated from the tracing simulation as the time between the particle origin (traced backward) and its subsequent absorption (traced forward) in the atmosphere (i.e. the tracing time $\tau$).
Results are displayed in Figure \ref{Figure1}, where the lifetime for the measured sample is shown as a function of
the AACGM latitude $\Lambda$ (top-left panel), and of the product $R\cdot L^{2}$ (top-right panel), where $R$ is the particle rigidity and $L$ is the McIlwain's parameter \citep{McIlwain} in units of Earth's radii (R$_{E}$); note that $R\cdot L^{2}$ $\simeq$ 14.3 GV R$_{E}^{2}$ corresponds to the St\"{o}rmer vertical cutoff \citep{STORMER,SHEA} for the PAMELA epoch.
Distributions as a function of kinetic energy are also reported in bottom panels, for magnetic latitudes $|\Lambda|<$35 deg (left) and $|\Lambda|>$35 deg (right) respectively.

The several albedo components can be easily distinguished.
QT proton lifetimes typically amount to $\tau \sim$ 0.3 $\div$ 30 s, while the upper bands ($\tau$ up to $\sim$ 2 min) correspond to particles performing more than a revolution around the Earth. Since the QT proton lifetime is of the order of a half drift period, which scales with $\sim 1/\gamma\beta^{2}$ \citep{Walt}, lifetime and energy have an approximately inverse proportional relation.

Conversely, the UT$_{S}$ proton lifetime ranges from some fraction of $s$ up to a few $s$ (depending on latitude) and it is shorter than the typical bounce period, which scales with $1/\beta$, resulting in a weaker dependency on energy.
In particular, their distribution is given by two superimposed bands, corresponding to particles crossing the magnetic equator once and twice, respectively. In addition, a population of very-short-lived ($<30$ $ms$)
protons emerges at high energies ($0.8\div8$ GeV) and low latitudes ($|\Lambda|<$ 10 deg), with a tracing time increasing with energy: because of the large gyro-radius with respect to the magnetic field curvature, such particles have no stable bounce motion ($\omega_{bounce}\sim\omega_{gyro}$, large $\omega_{drift}$ values).

Non-adiabatic or large gyro-radius effects cause the breakdown of (quasi-) trapping conditions for $R\cdot L^{2}$ $\gtrsim$ 4 GV R$_{E}^{2}$.
Such a geomagnetic region is populated by UT$_{L}$ protons (see top-right panel), characterized by irregular trajectories with no periodicity; consequently, their lifetime distributions are much broader, extending up to a few tens of min (pseudo-trapped motion).
Such a component can be energetic enough to populate the penumbra region ($R\cdot L^{2}$ $\gtrsim$ 10 GV R$_{E}^{2}$), where particles of both atmospheric and galactic origin are present.

\begin{figure*}[h!]
\centering
\includegraphics[width=6.6in]{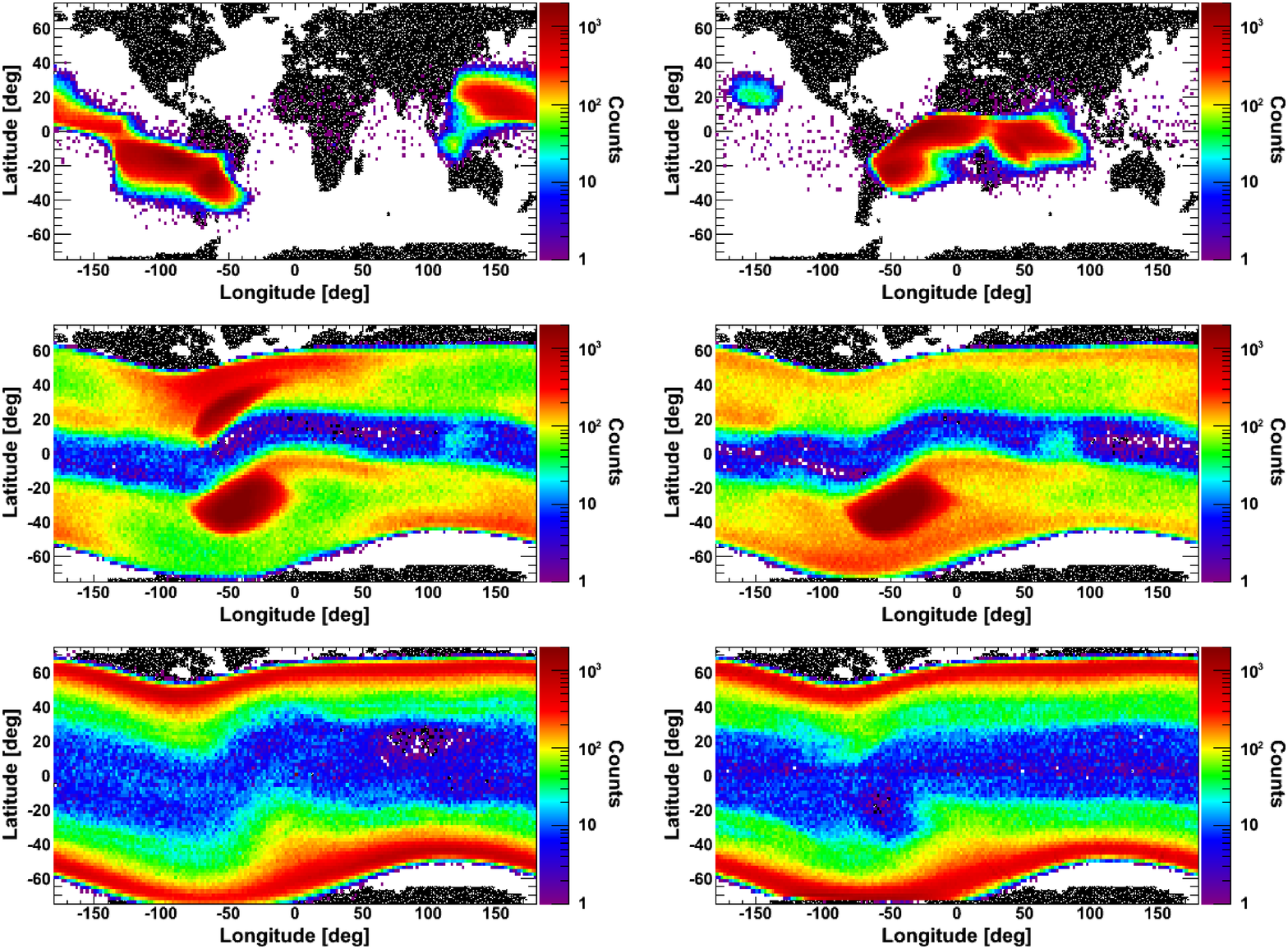}
\caption{Count distributions of the production (left panels) and absorption (right panels) points on the atmosphere (40 km) for quasi-trapped (top), precipitating (middle) and pseudo-trapped (bottom) protons as a function of the geographic coordinates.}
\label{Figure2}
\end{figure*}

\subsection{Production and Absorption Points on the Atmosphere}
The production and absorption points on the atmosphere are shown in Figure \ref{Figure2}, respectively in left and right panels. In the case of QT protons (upper panels), the former are located
in a region extending westward from the SAA, where the Earth's magnetic field is at its minimum due to asymmetries 
between terrestrial magnetic dipole and rotational axes
(tilt $\sim$10 deg, offset $\sim$500 km).
While drifting from the SAA, protons encounter stronger magnetic fields and the altitude of their mirror points increases, until they reach again weaker magnetic field regions; then their mirroring altitude decreases and finally they are absorbed by the atmosphere, mainly on the region on the East side of the SAA. Both source and sink points are located in two regions, in the southern and in the northern magnetic hemisphere respectively, as a consequence of the multipole moment of the Earth's magnetic field \citep{Huang}.

Conversely, since UT$_{S}$ protons (middle panels) are created and absorbed by the atmosphere in a very short time, their production and absorption points are located near the detection point, uniformly populating the whole geomagnetic region explored by PAMELA. In particular, UT$_{S}$ sink points have a peak in the SAA, while origin points have two peaks: the former in the SAA, the latter in the northern magnetic region corresponding to southern mirror points in the SAA. Particles produced in the SAA can undergo a reflection in the upper hemisphere (where mirror point altitude is higher) before being re-absorbed, and correspond to the upper band ($\tau$ larger than a half bounce period) in the UT$_{S}$ lifetime distributions (see Figure \ref{Figure1}).

Similar to UT$_{S}$ protons, UT$_{L}$ distributions (bottom panels) spread over all longitudes. In addition, their absorption points have a minimum in the SAA, while source distribution has a minimum in
the geomagnetic region opposite to the SAA (around $\sim$20$^{\circ}$ N, $\sim$90$^{\circ}$ E).
It should be noted that reported distributions were evaluated at a fixed altitude of 40 km, used as a limit for the absorbing atmosphere, while the actual source and sink positions are expected to be broader.

\begin{figure*}[ht]
\centering
\includegraphics[width=6.6in]{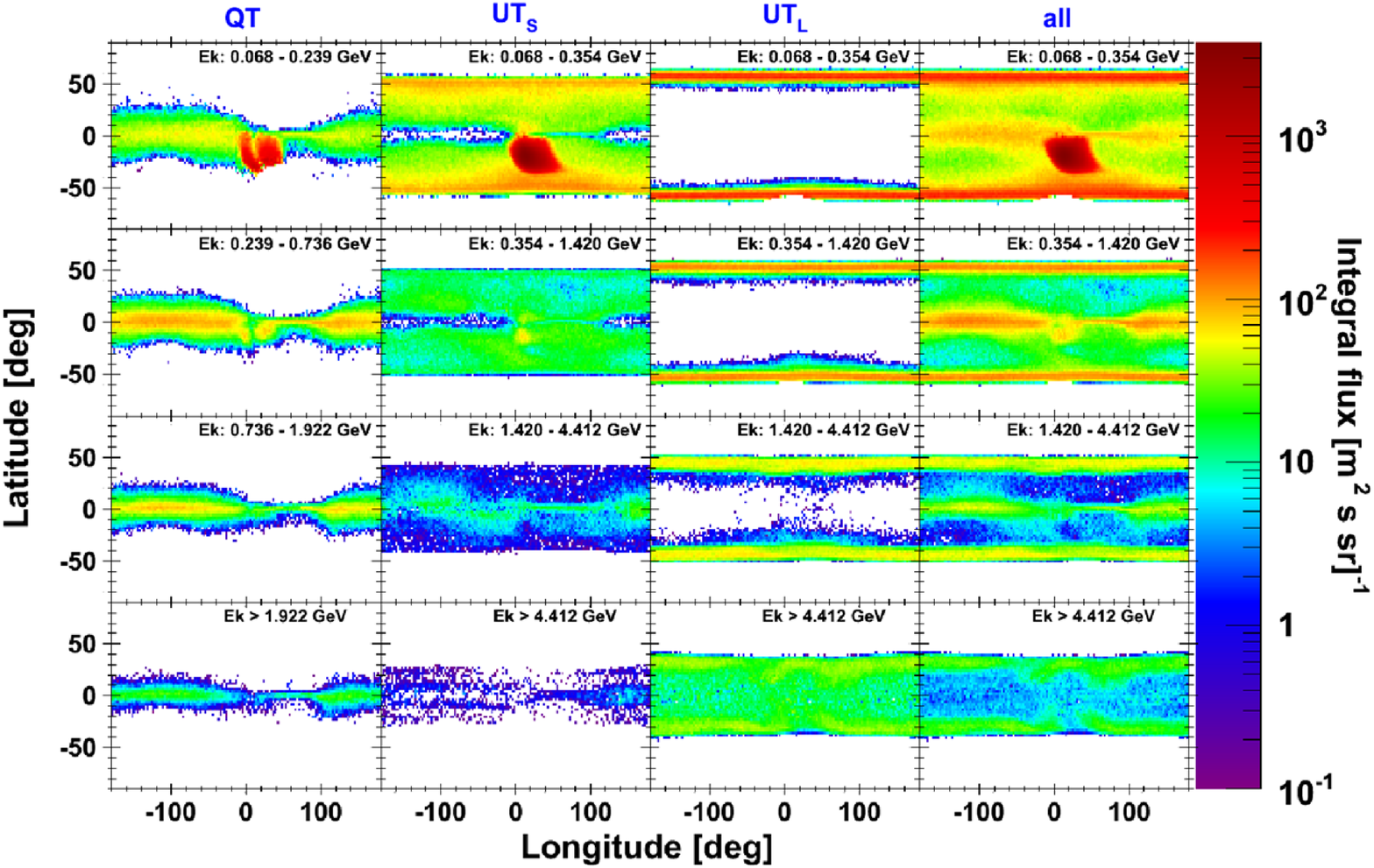}
\caption{Proton integral fluxes (m$^{-2}$ s$^{-1}$ sr$^{-1}$) as a function of AACGM longitude and latitude, for different energy bins. Results for the several albedo populations are reported (from left to right): quasi-trapped (QT), precipitating (UT$_{S}$) and pseudo-trapped (UT$_{L}$) protons, along with the total sample. Note that a different energy binning is used for QT protons (left column).}
\label{Figure3}
\end{figure*}

\section{Albedo Proton Fluxes}
Fluxes were derived for bins of AA\-CGM longitude, latitude and proton energy, by assuming an approximately isotropic particle distribution. Integral flux maps (m$^{-2}$ s$^{-1}$ sr$^{-1}$) in geomagnetic coordinates are displayed in Figure \ref{Figure3} for different energy bins. Results for the different populations are reported.
The flux dependency on longitude
is quite complex, reflecting the asymmetries between the magnetic and geographic axes, and it is significantly influenced by the presence of non-dipolar components of the geomagnetic field. In addition, at PAMELA energies the particle gyro-radius is quite large, affecting the portion of fluxes which can be investigated by PAMELA, so that measured distributions have different shapes for different energies.
Low energy QT and UT$_{S}$ distributions display a peak near the SAA (longitude $\sim$ 0 $\div$ 50 deg), where the Earth's magnetic field is at its minimum. In addition, we found that measured energy spectra become harder while approaching to the geomagnetic region opposite to the SAA, where the geomagnetic field has a local maximum. Fluxes were averaged over the pitch angle and
the altitude ranges covered by PAMELA.
An average flux increase of a factor $\sim$ 1.5 was observed between lowest and highest explored altitudes.

\begin{figure*}[ht]
\centering
\noindent
\includegraphics[width=6.6in]{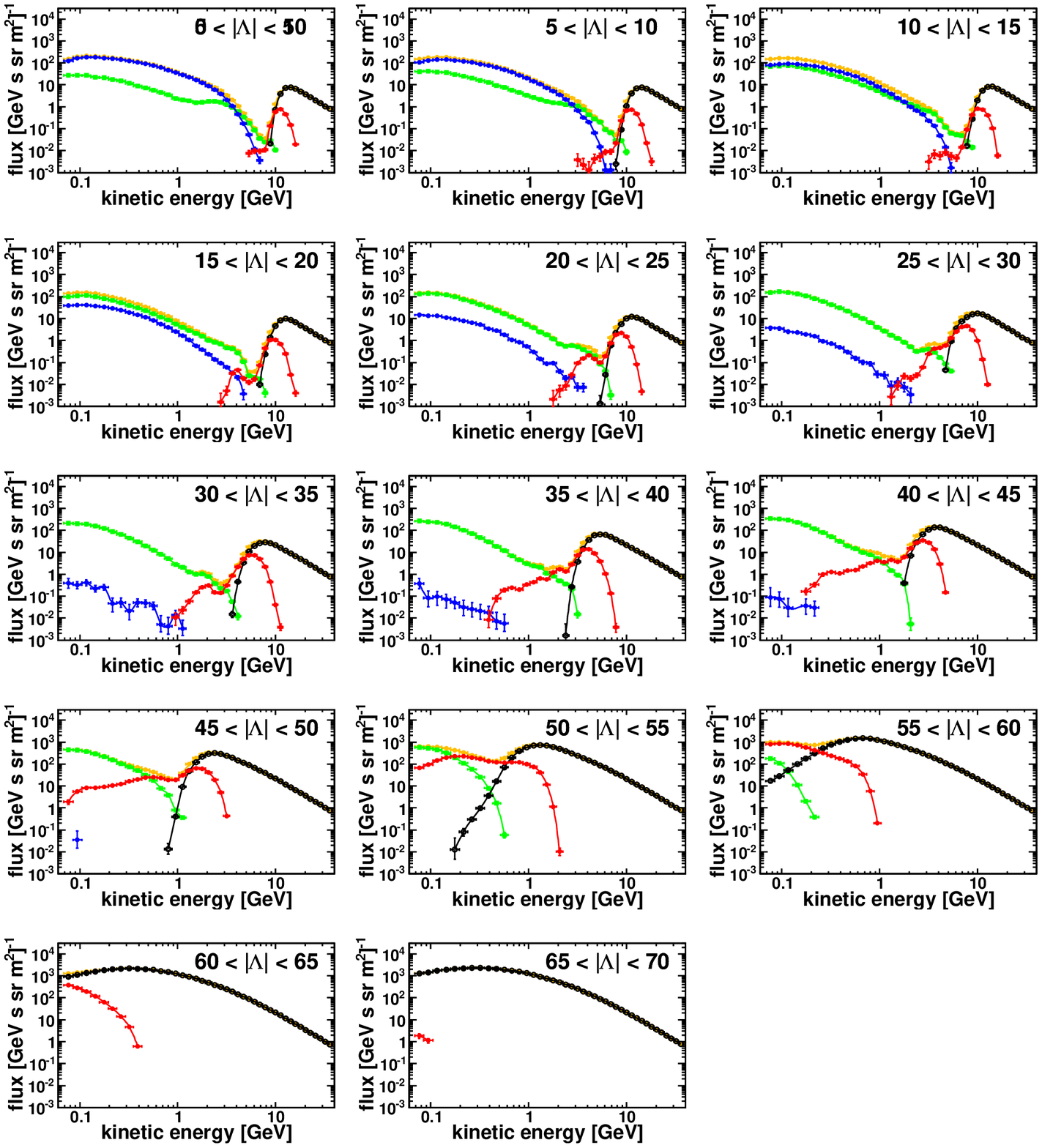}
\caption{Differential energy spectra (GeV$^{-1}$ m$^{-2}$ s$^{-1}$ sr$^{-1}$) outside the SAA region measured for different bins of AACGM latitude $|\Lambda|$ (see the labels). Results for the several proton populations are shown: quasi-trapped (QT, blue circle), precipitating (UT$_{S}$, green squares), pseudo-trapped (UT$_{L}$, red triangles) and galactic (black empty circles); the underlying orange points denote the total proton flux. Lines are to guide the eye.}
\label{Figure4}
\end{figure*}

\subsection{Energy Spectra}
Energy spectra were evaluated inside and outside the SAA, by requiring local magnetic field intensities lower and higher than 0.23 G, respectively. Figure \ref{Figure4} shows the spectra outside the SAA measured at different AACGM latitudes $|\Lambda|$. Fluxes were averaged over longitudes. Reported error bars account for statistical and systematic uncertainties, and are smaller than the plotted points in most cases. The several proton components (QT, UT$_{S}$ and UT$_{L}$) are denoted with different colors (blue, green and red, respectively); galactic proton spectra (black points) are also reported for a comparison;
finally, the underlying orange points denote the total (albedo+galactic) proton spectra.

\begin{figure*}[!ht]
\centering
\includegraphics[width=6.6in]{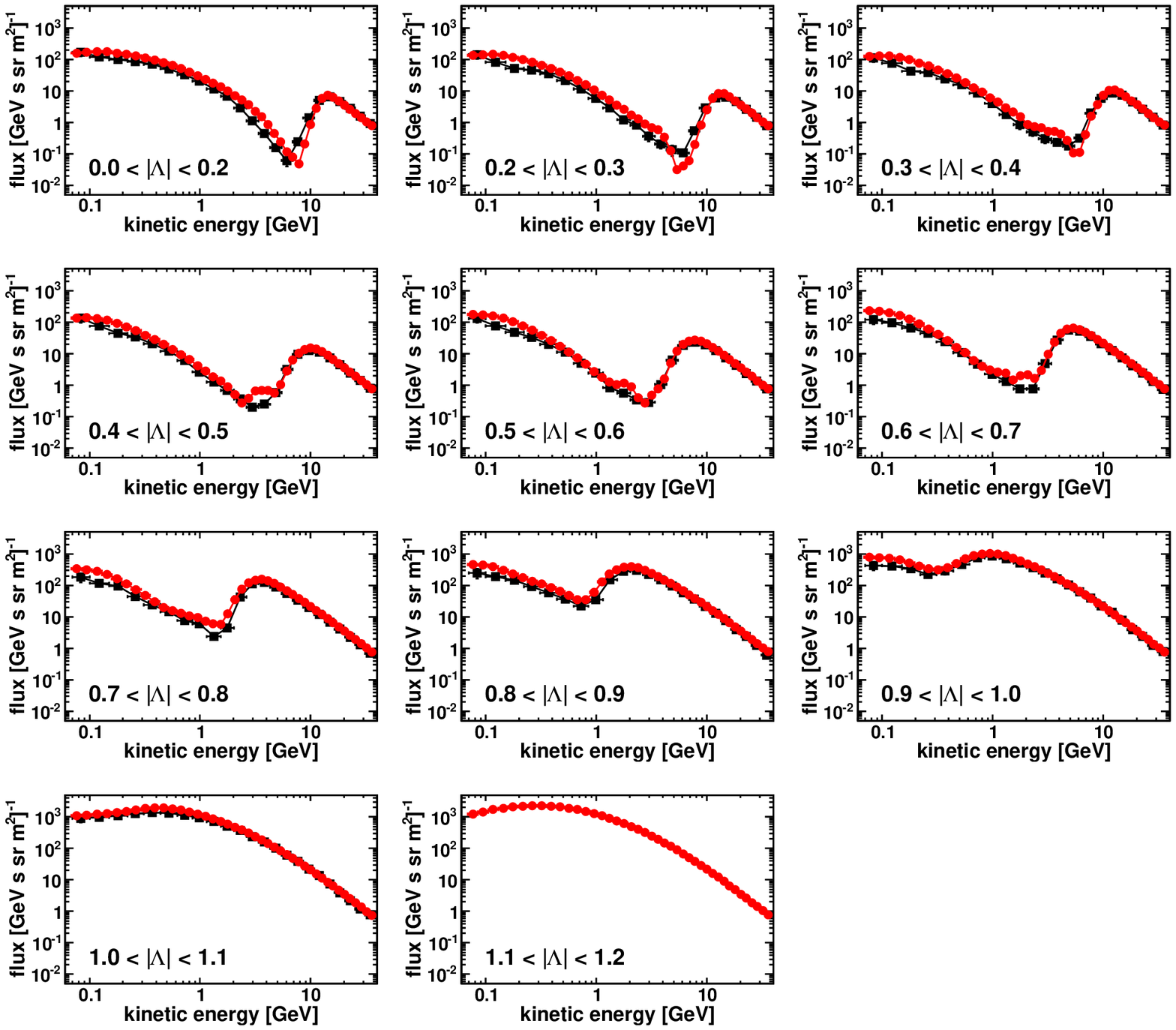}
\caption{Total proton spectra, including both galactic and albedo components, for different bins of geomagnetic latitude $|\Lambda|$ (radians). For comparison, AMS-01 \citep{AMS01b} results (black points, $|\Lambda|$ $<$ 1.1 rad) are also reported. Lines are to guide the eye.}
\label{Figure5}
\end{figure*}

Fluxes of QT protons are limited to low latitudes and to energies below $\sim$ 8 GeV; they smoothly decrease with increasing latitude and energy.
Conversely, UT$_{S}$ distribution spreads to higher latitudes, with energetic spectra extending up to $\sim$10 GeV.
Their flux is suppressed at the equator, since corresponding magnetic field lines typically do not reach PAMELA altitudes.
Both QT and UT$_{S}$ components have a low energy peak in the SAA, while spectra measured outside the SAA are harder.

\begin{figure*}[ht]
\centering
\includegraphics[width=6.8in]{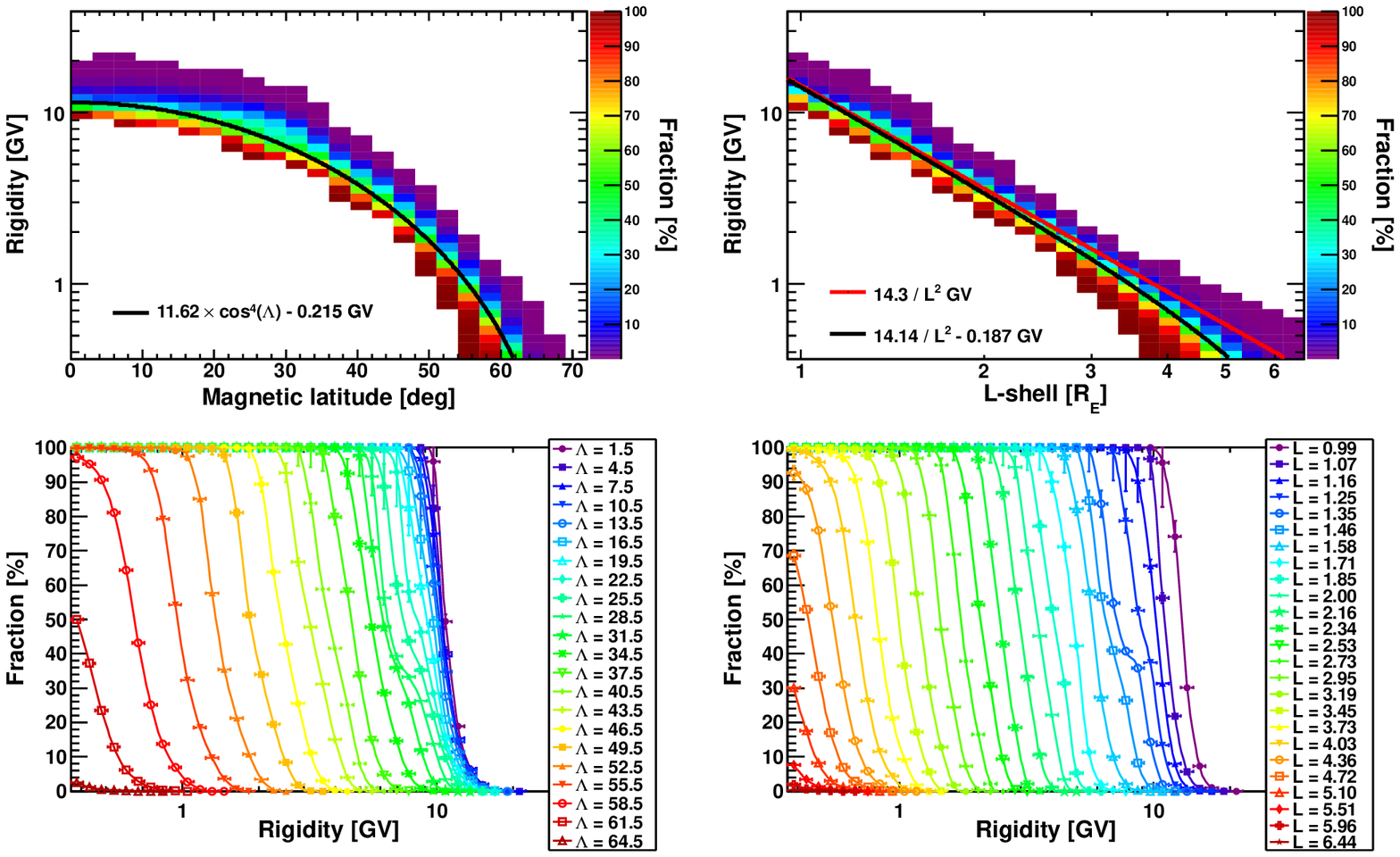}
\caption{Top panels: fraction of albedo protons in the penumbra region, as a function of particle rigidity and AACGM latitude $|\Lambda|$ (left) and McIlwain's $L$-shell (right); the black curves are a fit of points with equal percentages of galactic and albedo protons, while the red curve denotes the St\"{o}rmer vertical cutoff for the PAMELA epoch. Bottom panels: corresponding rigidity profiles, for different values of $|\Lambda|$ (left) and $L$ (right); values at bin center are reported in labels. Lines are to guide the eye.}
\label{Figure6}
\end{figure*}

Finally, the UT$_{L}$ protons concentrate at the highest latitudes and energies (up to $\sim$ 20 GeV),
with a peak in the penumbra.
Such a feature is due to large gyro-radius ($10^{2}\div10^{3}$ km) effects: particles with increasing rigidity come from a more and more extended region causing fluxes to increase, until trajectory curvature becomes too large and they escape from the magnetosphere, and fluxes decrease again \citep{BOBIK}. Corresponding pitch angle distributions are much broader, extending to highest values.
Data supporting Figure \ref{Figure4} are available in the Supporting Information Tables S1--S4.


Figure \ref{Figure5} reports the comparison between total proton spectra outside the SAA ($B$ $>$ 0.26 G), including both galactic and albedo components, measured by PAMELA (red) and AMS-01 (black), for different bins of geomagnetic latitude $|\Lambda|$.
Note that AMS fluxes were evaluated (using CGM coordinates \citep{BREKKE}) at altitudes of 350 $\div$ 390 km and are limited to geographic latitudes lower than $\sim$ 52 deg.
Latitudes are reported in radians, for consistency with AMS published results.
Despite good qualitative agreement, significant differences can be observed. AMS spectra results up to 50\% lower than PAMELA ones, especially at low energies and latitudes.
On the other hand, AMS fluxes are a factor 2 $\div$ 3 larger than PAMELA ones near the penumbra.
In addition, PAMELA spectra show some structures near the dip region before the penumbral peak, especially at middle-low latitudes, which are absent in AMS data.

\subsection{Penumbra Protons}\label{Penumbra protons}

Features of the penumbra region can be more deeply investigated in Figure \ref{Figure6}, where the fraction of albedo protons is displayed as a function of particle rigidity and AACGM latitude (left panels); for a comparison, distributions as a function of McIlwain's $L$-shell are also shown (right panels). Bottom panels report corresponding rigidity projections, including statistical uncertainties on measured distributions.
The penumbra was identified as the region where PAMELA detected both albedo and galactic proton trajectories.

The black curves denote
a fit of points with an equal percentage of the two components:
\begin{equation}\nonumber
\centering
R_{50\%}^{fit}(\Lambda) \simeq 11.62\cdot cos^{4}\Lambda - 0.215 \hspace{0.3cm} \text{GV}
\end{equation}
and
\begin{equation}\nonumber
R_{50\%}^{fit}(L) \simeq 14.14/L^{2}-0.187 \hspace{0.3cm} \text{GV}
\end{equation}
for magnetic latitude $\Lambda$ and $L$-shell distributions, respectively.
For a comparison, the red curve indicates the St\"{o}rmer vertical cutoff evaluated for the PAMELA epoch: $R_{SV}(L) \simeq 14.3/L^{2}$ GV.
The fitting functions $R_{50\%}^{fit}$ are similar to the one used by \citep{Ogliore} to reproduce cutoff observations made by the MAST instrument on the SAMPEX satellite, and they account for magnetospheric effects at high latitudes.
However, because of the non-dipole terms in the geomagnetic field \citep{SMART}, the fit results can vary up to a factor 10\% at equatorial and mid-latitude locations.

\begin{figure}[t]
\centering
\includegraphics[width=3.3in]{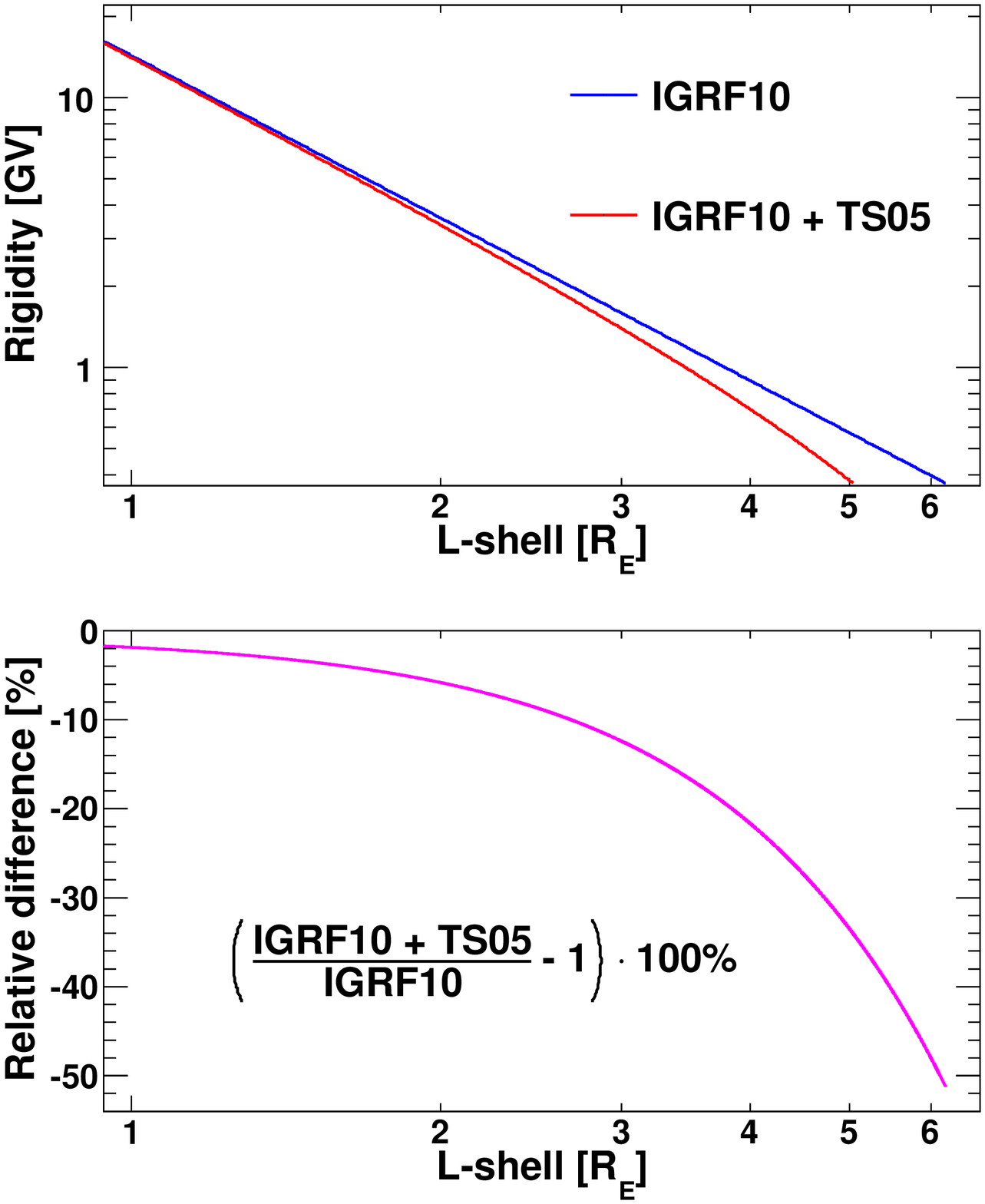}
\caption{Top panel: comparison between fit values $R_{50\%}^{fit}(L)$ obtained with (red) and without (blue) the inclusion of external geomagnetic field sources, as a function of $L$-shell. Bottom panel: corresponding relative difference in percentage.}
\label{Figure7}
\end{figure}

The usage of a realistic external geomagnetic field model has a significant impact on this study influencing the reconstructed fraction of albedo protons, as highlighted in Figure \ref{Figure7}.
The top panel compares the estimated $R_{50\%}^{fit}(L)$ values derived from PAMELA data with and without the inclusion of external field sour\-ces, corresponding to the red (IGRF10 + TS05) and the blue curve (IGRF10) respectively. In the latter case, we found that fit results ($R_{50\%}^{fit}(L)$ $\simeq$ 14.22/L$^{2}$ GV) are in a very good agreement with the standard cutoff relation $R_{SV}(L)$ \citep{SHEA1985} at all $L$-shells.
On the other hand, significant differences (up to a factor $\sim$ 2)
can be noted between the IGRF10 model and the combined
model configuration
at high $L$, as shown in the bottom panel.
Similar results were obtained for the $\Lambda$ distribution.

\begin{figure}[t]
\centering
\includegraphics[width=3.3in]{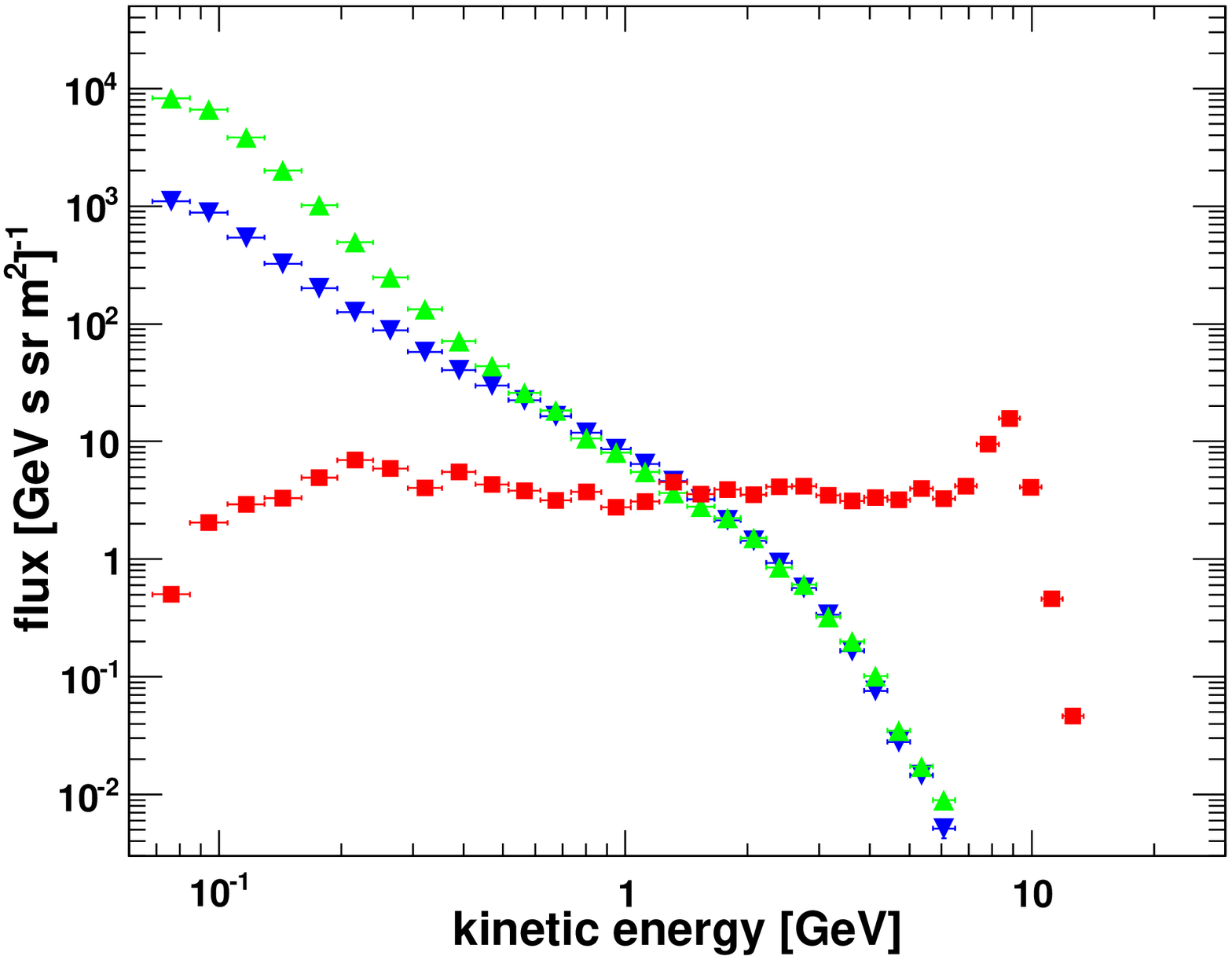}
\caption{Mean differential energy spectra (GeV$^{-1}$ m$^{-2}$ s$^{-1}$ sr$^{-1}$) in the SAA region ($B$ $<$ 0.23 G). Results for the different proton populations are shown: quasi-trapped (QT, blue), precipitating (UT$_{S}$, green) and pseudo-trapped (UT$_{L}$, red).}
\label{Figure8}
\end{figure}

\subsection{Albedo Fluxes in the SAA}\label{Albedo fluxes in the SAA}
Figure \ref{Figure8} shows the average energy spectra of QT, UT$_{S}$ and UT$_{L}$ protons in the SAA.
Measured
spectra are softer than those in other magnetic regions,
with a significant flux increase at low energies.
In case of QT and UT$_{S}$ protons, an additional contribution is possibly expected from the decay of free albedo neutrons, which constitutes the main source of geomagnetically trapped protons in radiation belts \citep{Singer,Farley}.
The UT$_{L}$ component emerges at high energies ($\gtrsim$ 1.5 GeV), concentrating near the local geomagnetic cutoff.

\section{Conclusions}\label{Conclusions}
PAMELA measurements of energetic ($>$ 70 MeV) albedo proton fluxes at low Earth orbits (350 $\div$ 610 km) have been presented. The detected sample, corresponding to data acquired by PAMELA between 2006 July and 2009 September, was analyzed and classified on the basis of a realistic
simulation of particle trajectories in the Earth's magnetosphere
into quasi-trapped and un-trapped protons: the former consists of relatively long-lived protons, detected in the near equatorial region, with trajectories similar to those of stably-trapped protons from the inner radiation belt; the latter was found to spread over all explored latitudes, including short-lived (precipitating) protons together with a long-lived (pseudo-trapped) component constituted by particles with rigidities near the local geomagnetic cutoff and
characterized by a chaotic motion (non-adiabatic trajectories). Fluxes were mapped by using the
AACGM coordinates to provide a more realistic description at higher latitudes.

PAMELA results significantly enhance the characterization of high energy albedo proton populations
in a wide geomagnetic region,
including the SAA and the penumbra,
enabling a more accurate and complete view of atmospheric and geomagnetic effects
on the CR transport
near the Earth.

\section*{Acknowledgements}
We acknowledge support from The Italian Space Agen\-cy (ASI), Deutsches Zentrum f\"{u}r Luftund Raumfahrt (DLR), The Swedi\-sh National Space Board, The Swedi\-sh Research Council, The Russian Space Agency (Ros\-cosmos) and The Russian Foundation for Basic Research.
We gratefully thank R. Selesnick, M. Honda and J. F. Cooper for helpful discussions, D. F. Smart and M. A. Shea for the assistance with their trajectory code, and N. Tsyganenko for the support in the use of the TS05 model.


\end{document}